\documentclass[%
 aip,
 amsmath,amssymb,
reprint,%
]{revtex4-1}

\usepackage{graphicx}
\usepackage{dcolumn}
\usepackage{bm}

\usepackage[utf8]{inputenc}
\usepackage[T1]{fontenc}
\usepackage{mathptmx}
\usepackage{etoolbox}
\usepackage{color}
\usepackage[mathlines]{lineno}

\makeatletter
\def\@email#1#2{%
 \endgroup
 \patchcmd{\titleblock@produce}
  {\frontmatter@RRAPformat}
  {\frontmatter@RRAPformat{\produce@RRAP{*#1\href{mailto:#2}{#2}}}\frontmatter@RRAPformat}
  {}{}
}%
\makeatother
\begin{document}

\preprint{AIP/123-QED}

\title[Study of Interaction and Complete Merging of Binary Cyclones Using Complex Networks]{ Study of Interaction and Complete Merging of Binary Cyclones Using Complex Networks}
\author{Somnath De}
\email{somnathde.mec@gmail.com}
\affiliation{Department of Aerospace Engineering, Indian Institute of Technology  Madras, 600036, India}%
 
\author{Shraddha Gupta}%
\email{shraddha.gupta@pik-potsdam.de}
\affiliation{Potsdam Institute for Climate Impact Research (PIK) -- Member of the Leibniz
Association, Telegrafenberg A56, Potsdam, 14473, Germany}%
\affiliation{Department of Physics, Humboldt University at Berlin, Newtonstraße 15, Berlin, 12489, Germany}%

\author{Vishnu R Unni}
\affiliation{Department of Mechanical and Aerospace Engineering, Indian Institute of Technology Hyderabad, 502284, India}%

\author{Rewanth Ravindran}
\affiliation{Department of Aerospace Engineering, Indian Institute of Technology  Madras, 600036, India}%

\author{Praveen Kasthuri}
\affiliation{Department of Aerospace Engineering, Indian Institute of Technology  Madras, 600036, India}%

\author{Norbert Marwan}
\affiliation{Potsdam Institute for Climate Impact Research (PIK) -- Member of the Leibniz
Association, Telegrafenberg A56, Potsdam, 14473, Germany}%
\affiliation{Institute for Geosciences, University of Potsdam, Karl-Liebknecht-Straße 24-25, Potsdam, 14476, Germany}

\author{J\"urgen Kurths}
\affiliation{Potsdam Institute for Climate Impact Research (PIK) -- Member of the Leibniz
Association, Telegrafenberg A56, Potsdam, 14473, Germany}%
\affiliation{Department of Physics, Humboldt University at Berlin, Newtonstraße 15, Berlin, 12489, Germany}%

\author{R. I. Sujith}%
\affiliation{Department of Aerospace Engineering, Indian Institute of Technology  Madras, 600036, India}%

\date{\today}

\begin{abstract}

Cyclones are amongst the most hazardous extreme weather events on Earth. When two co-rotating cyclones come in close proximity, a possibility of complete merger (CM) arises due to their interactions. However, identifying the transitions in the interaction of binary cyclones and predicting the merger is challenging for weather forecasters. In the present study, we suggest an innovative approach to understand the evolving vortical interactions between the cyclones during two such CM events using time-evolving induced velocity based unweighted directed networks. We find that network-based indicators, namely, in-degree and out-degree, can quantify the changes during the interaction between two cyclones and are better candidates than the traditionally used separation distance to classify the interaction stages before a CM. The network indicators also help to identify the dominating cyclone during the period of interaction and quantify the variation of the strength of the dominating and merged cyclones. Finally, we show that the network measures also provide an early indication of the CM event well before its occurrence.
\end{abstract}


\maketitle

\begin{quotation}
In some active cyclone basins, more than one cyclone can be formed concurrently. Consequently,  two or more cyclones can come in close spatial proximity and interact with each other. Such an interaction may lead to many possibilities, like weakening of both cyclones, sudden alteration in their tracks, re-strengthening of one of the cyclones due to vorticity interaction, and very rarely birth of a more intense, longer-lived cyclone due to complete merging between them. Binary interaction between cyclones has not been fully understood and is a major challenge for weather forecasters. This often leads to inaccurate predictions, increasing the risk of human life and property due to unpreparedness. Most previous observations and model-based investigations have used the separation distance between the cyclones to classify the stages of binary interaction leading to merging and to predict their merger. Although the separation distance can be both the cause and the effect of mutual interaction between the cyclones, but it may not be the most suitable parameter to classify the binary interaction stages. In this study, we use a novel approach based on complex networks. We analyze the vortical interactions in the spatial domain by constructing the time-evolving induced velocity network at each time instant. Using two prominent examples of complete merger events, namely, the Seroja-Odette and Noru-Kulap interactions in the northern and southern hemispheres, respectively, we show that network-based measures are better and more informative in classifying the binary interaction stages.

\end{quotation}

\section{Introduction}\label{sec1}

Cyclones are organized non-frontal synoptic convective vortical systems that are formed over tropical or subtropical waters. Essentially, they are characterized by a low pressure center \cite{fink1998tropical} that produces strong surface wind circulation. When a cyclone makes landfall, torrential rains and the accompanying strong winds impart severe widespread damage to land infrastructure, disrupting human lives and even resulting in numerous casualties. The massive destruction caused by severe cyclones in recent years has raised serious concerns that these extreme weather events may be among the possible consequences of human-induced climate change. Due to global warming, sea surface temperatures are rising and the maximum capacity of the atmosphere to hold water vapour has also increased. A number of studies~\cite{knutson2019tropical} have indicated that anthropogenic global warming is likely to cause an increase in the intensity of cyclones, higher precipitation rates, and elevated storm surge risks. Tropical cyclones may also intensify more rapidly, have slower translation speeds, and occur at higher latitudes~\cite{knutson2019tropical}. Therefore, understanding the behavior of cyclones is of utmost interest to weather forecasters and policymakers. 

In some very active cyclone basins, such as the Northwestern Pacific and Atlantic, multiple cyclone systems can be formed simultaneously. Although rare, two cyclones can come within a close proximity and interact, beginning an intense dance about their common center, which may lead to the strengthening of the cyclones, sudden track changes, or even completely merging into one cyclone. Such an interaction of binary cyclones was first reported by Okada \cite{okada1907}. According to his observation, cyclones tend to come closer and intensify if they spin in the same direction while they tend to separate if they rotate in the opposite direction. Later, Fujiwhara made similar deductions on the amalgamation of cyclones through laboratory experiments and geophysical observations \cite{fujiwhara1923growth, fujiwhara1931short,fujiwhara1921natural}. Subsequently, this binary cyclone interaction, when multiple cyclone make a close pass, came to be known as the \textquotedblleft\textit{Fujiwhara effect}\textquotedblright. Thereafter, a number of weather events have been recorded where one cyclone has been observed to interact with another cyclone within close proximity \cite{liou2016generalized,liu2014analysis,hoover1961relative}. 

The Fujiwhara interaction often alters the tracks of the cyclones, making them difficult to forecast. Inaccuracies in predicting cyclone tracks increase the threat to life and property due to unpreparedness caused by misinformation and the lack of early warning. For instance, unforeseen heavy rainfall occurred in Taiwan, and the same region of the Luzan Island of the Philippines experienced landfall of typhoon Parma thrice due to its interaction with another typhoon Melor in October 2009, causing significant fatalities and economical losses~\cite{liou2020interactions}. In most cases, the Fujiwhara effect weakens both cyclones as the winds involved with cyclones in the same hemisphere during the interaction tend to blow opposite to each other. However, very rarely, the binary interaction may lead to a re-strengthening of the cyclone, as in the case of Category 3, severe tropical cyclone Seroja in April 2021 due to its complete merger with Odette~\cite{Seroja}. Interaction of a cyclone with other cyclonic vortices may also prolong its life span, e.g., the Super Typhoon Noru in July 2017 lasted for 19 days due to its successive dual vortex direct and indirect interactions with typhoons Kulap, Haiting and Nesat~\cite{liou2019consecutive}. Till date, it has not been possible to fully understand and incorporate the Fujiwhara effect in numerical weather prediction models to improve cyclone forecasts. Hence, it is highly essential to study such cases of binary cyclone interaction to deepen our understanding. 

Generally, based on a circulation-based vortex pair interaction, the interaction of cyclones is classified into five categories~\cite{prieto2003classification}, which are: (a) partial straining out, (b) complete straining out, (c) partial merger, (d) complete merger, and (e) elastic interaction. (a) and (c) signify the partial deformation of the interacting pair, while complete deformation of one of the interacting vortices can be found in (b) and (d). In (e), each interacting vortex survives, although its direction of motion changes. Among them, a complete merger (CM) of two cyclones is of great interest to the climate community because it is one of the most complicated interactions in the context of the transfer of energy and vorticity across the turbulent flow scales ~\cite{leweke2016dynamics}. Earlier studies \cite{leweke2016dynamics,meunier2005physics,cerretelli2003physical} based on theoretical calculations have shown that the transfer of fluid from one layer to others in a system of co-rotating vortices was the reason for their merging. However, such an inter-layer fluid exchange is not confirmed in real-world binary cyclone interaction. 

As a result, forecasting cyclone tracks when two low-pressure systems are in close spatial proximity is a challenging task. One of the important factors related to the error in the forecasts of the cyclone tracks is the presence of another low-pressure system in close spatial proximity~\cite{brand1970interaction,ForecastErrorBinaryTC,ForecastErrorBinaryTC2,JTWC_Forecasterror}.
Several studies \cite{hoover1961relative,brand1970interaction,dong1983relative,lander1993interaction} based on observational data,
found that although most mutual interactions close to the intertropical convergence zone (ITCZ) in the North Pacific agree with the Fujiwhara expectations, there were some notable exceptions, especially in the North Atlantic. Moreover, Lander and Holland \cite{lander1993interaction} 
in their detailed analysis on interacting cyclonic vortices in the western North Pacific found that the classical Fujiwhara model of CM is seldomly followed. They reported that the presence of large-scale clockwise circulation patterns masks the Fujiwhara effect, sometimes even at separation distances where the Fujiwhara forces are quite strong. Further, large-scale circulation due to the presence of subtropical high or monsoon depression~\cite{Forecasterror_TCmonsoon,lander1993interaction,ritchie1993interaction,TropicalCyclonesinMonsoonGyres} and the presence of multiple weak cyclonically-rotating meso-vortices~\cite{JTWC_Forecasterror} pose significant challenges towards cyclone track forecasts. Therefore, weighing the impact of binary interaction on cyclone track and intensity is essential to cyclone forecasters.

Several numerical and analytical studies on the interactions of binary cyclones, have been attempted in an effort to understand both two-dimensional \cite{liu2014analysis,leweke2016dynamics,liou2016generalized,liou2020interactions} and three-dimensional dynamics \cite{wei1983numerical, demaria1984comments} of the CM phenomena. Most of them underlined the significant role of the separation distance in the interaction between binary cyclones. Chang \cite{wei1983numerical} 
showed an agreement with Fujiwhara's description of CM in the absence of large-scale circulations using a three-dimensional cyclone model. Their investigation also showed that the displacement of one of the interacting cyclones in the mutual rotation is proportional to the combined strength of the binary system but is inversely proportional to the size of the cyclone and to the square of the separation distance. On the contrary, their simulations using the non-divergent barotropic model in which the vortices interact by advection alone showed no signs of mutual attraction. However, DeMaria and Chan \cite{demaria1984comments}, later demonstrated that the mutual attraction can be explained using vorticity advection alone and is strongly dependent on the initial wind profile of the vortices. A number of studies~\cite{brand1970interaction,roberts1972topics,saffman1980equilibrium,overman1982evolution,dritschel1986nonlinear,meunier2002merging} 
found that merging occurs when the sizes of the vortex cores of co-rotating vortices increase beyond a critical fraction of the separation distance due to viscous diffusion. Further, several dissipative and convective stages \cite{cerretelli2003physical,josserand2007merging} are identified based on the separation in the vortex merging process. Such occurrence of a rapid merger following the approach of cyclone-scale vortices within a critical separation distance was reported from simulations~\cite{lander1993interaction,ritchie1993interaction,holland1993interaction} of a modified model of binary interaction. However, there have not been much investigations on the dynamics of CM based on observation or reanalysis data to compare with these model-based findings.

Despite the numerous studies on the shearing of cyclones when in close proximity
\cite{liou2020interactions, wu2003new,liu2015interactions}, the interaction of two neighboring cyclones before a CM is not well explored due to the paucity of the occurrence of such merging events in nature. To that end, in the current study, we select two recent binary cyclone systems -- Noru-Kulap (during 23$^{rd}$ - 26$^{th}$ July 2017)~\cite{li2021intensity,liou2019consecutive} occurring in the northern hemisphere, and Seroja-Odette (5$^{th}$ - 10$^{th}$ April 2021) in the southern hemisphere -- which engaged in a Fujiwhara interaction and exhibited a CM event. Following the interaction of the severe tropical cyclone Seroja with the tropical storm Odette, the CM event steered the merged cyclone southward towards Australia and further strengthened it, as mentioned earlier. Then, the merged cyclone made landfall on the west coastline of Western Australia as a Category 3 severe tropical cyclone causing significant damage. Its prolonged southward trajectory was highly unusual as cyclones of similar intensity have travelled so far south only 26 times in the past 5000 years~\cite{Seroja,nott20116000}. Similarly, the Category 4 Super Typhoon Noru, the third longest-lived cyclone on record in the Northwest Pacific Ocean, became the second most intense tropical cyclone of the Northwestern Pacific Ocean basin in 2017, due to Fujiwhara interaction with Kulap and indirect interactions with other cyclone systems~\cite{liou2019consecutive}. Noru brought torrential rainfall to southern and western parts of Japan that triggered widespread flooding and caused large economic losses~\cite{2017impact}. In view of the aforementioned discussion, we need an approach that can enable us to gain deep insights into the topological structure and dynamics of such a highly complex weather system.

In recent decades, complex networks theory has emerged as one of the most powerful tools in understanding the interactions between the different units of a complex system across various disciplines~\cite{strogatz2001exploring,newman2003structure,de2005complex,krioukov2012network,barabasi2013network,krishnan2021suppression}. Tsonis et al. \cite{tsonis2004architecture} first applied this theory to study climate, by considering the climate system to be represented by a grid of oscillators interacting with each other in a complex way, with each one representing climate variability of a particular location of the gridded spatiotemporal dataset. Since then, the network representation of spatiotemporal climate data has been very successfully applied to study different climate and weather phenomena \cite{donges2009complex,donges2009backbone,marwan2015complex,meng2017percolation,ekhtiari2019disentangling,stolbova2014topology,malik2012analysis,stolbova2016tipping,boers2013complex, boers2015extreme,boers2019complex,ludescher2021network,boers2021complex}.  

Recently Gupta et al.~\cite{gupta2021complex} used time-evolving complex networks of mean sea level pressure (MSLP) data to study cyclones in the North Indian Ocean and tropical North Atlantic Ocean basins. They demonstrated that network-based indicators can be used to characterize the topological evolution of the regional climate system during highly localized weather extremes which occur over short time scales such as cyclones and detect cyclone tracks, besides climate phenomena like monsoon and the El Ni\~no-Southern Oscillation (ENSO) with occur over seasonal or annual time scales.

In the present work, we study the vortical interactions between two cyclones in close proximity leading to a CM, under a novel framework based on time-evolving induced velocity based unweighted networks. The adoption of induced velocity network based on the Biot-Savart law has been successfully used to study the turbulent flow dynamics~\cite{taira2016network}. Here, we extend the methodology to investigate flow dynamics in cyclonic systems (refer Sec. \ref{subsec:2}). In contrast to the correlation-based networks~\cite{gupta2021complex,fan2021statistical} which depicts only statistical relationships, the induced velocity networks represent real physical links indicating the induction of velocity by one flow element on the others. By considering the instantaneous vorticity field as a directed spatiotemporal network, we compute network measures, such as the \textit{in-degree} and the \textit{out-degree}, which count the number of links going to and arising from a particular grid point (see Sec. \ref{subsec:3}). This enables us to track the changes in the interaction zone of the binary cyclone system at every instant, as they approach each other, instead of examining the whole lifespan of the cyclones as in Gupta et al.~\cite{gupta2021complex}. We find that changes in the in-degree reflect the strength of the mutual interaction between two cyclones, while those in the out-degree are indicative of the vortical interactions within a cyclone. Our results show that as the two cyclones approach each other, the transitions occur in the network topology, which can be used to classify the complete merging process into several interaction stages. Therefore, our analysis does not only help to characterize the evolution of the cyclone~\cite{gupta2021complex} but also quantifies better the mutual interaction when it is in the vicinity of another cyclone system compared to separation distance~\cite{brand1970interaction,roberts1972topics,saffman1980equilibrium} and gives an early indication of the CM.

The rest of the paper is organized as follows. In Sec. \ref{sec:2}, a detailed description is provided about the source of data and the method of the construction of the network, which is used in the present study. In Sec.~\ref{subsec:5}, we perform a spatial analysis based on network measures, such as in-degree and out-degree, to understand the temporal evolution of vortical interactions between the two converging cyclones.
In Sec.~\ref{subsec:6}, we find the transitions exhibited by the maximum of the in-degree and the out-degree of the time-evolving networks, which enable us to classify the stages of the mutual interaction and merging between two cyclones. Finally, the significant remarks from the study are summarized in Sec.~\ref{sec:4}. 

\section{Methodology}\label{sec:2}
\subsection{Reanalysis Dataset}\label{subsec:1}

In the present work, we use the relative vorticity ($\omega$) data obtained from the state-of-the-art ERA5 reanalysis dataset \cite{hersbach2020era5} to understand the interaction dynamics between two co-rotating cyclones. Relative vorticity is defined as the rotation of air about a vertical axis, relative to a fixed point on the Earth's surface and calculated as $\omega$ = ${\frac{\partial v}{\partial x}}-{\frac{\partial u}{\partial y}}$, where, $u$ and $v$ corresponds to the velocity along $x$ (longitude) and $y$ (latitude), respectively. 

\begin{figure*}[ht!]%
\centering
\includegraphics[width=1.0\textwidth]{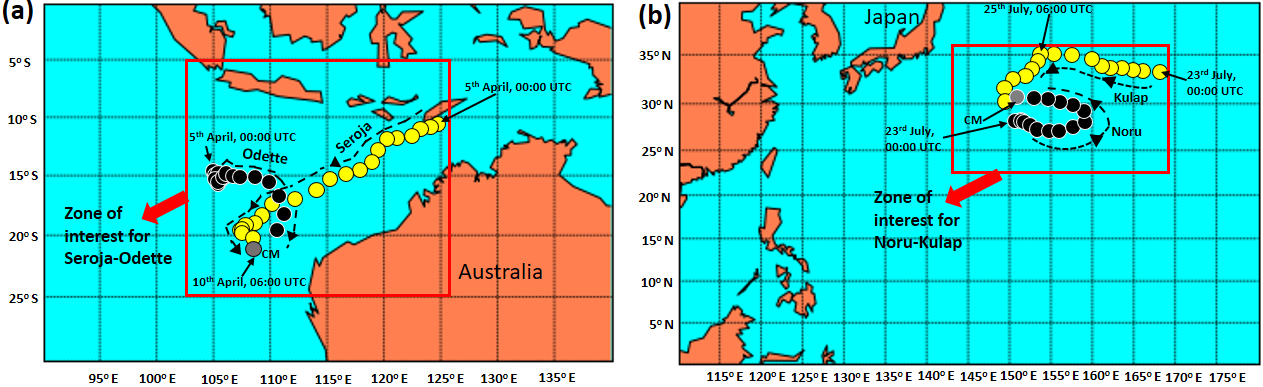}
\caption{The regions of interest for the (a) Seroja-Odette  and (b) Noru-Kulap interactions for which networks are constructed are shown here, respectively. For the Seroja-Odette interaction, the region of interest extends from 102$^o$ E to 125.5$^o$ E and from 5$^o$ S to 25$^o$ S. In case of the Noru-Kulap interaction, the area extends from 143$^o$ E to 169.5$^o$ E and from 23.5$^o$ N to 35.5$^o$ N. The trajectories of the cyclones at time steps of 6 hours are also shown to visually justify the selection of the spatial domain.}
\label{fig_window}
\end{figure*}

Environmental factors, especially large-scale relative vorticity at lower atmospheric levels (500-850 hPa), significantly affect cyclones \cite{chen2014variations,wu2020relative,gray1968global} and influences their relative motion in the presence of another cyclone \cite{dong1983relative}. However, most previous studies  \cite{wei1983numerical,wu2003new,liou2016generalized,liou2020interactions} on binary cyclone interaction found it difficult to correctly incorporate these large-scale environmental circulation in cyclone models, leading to erroneous prediction of cyclone tracks. Further, relative vorticity can represent the local features in the evolution of cyclones. In contrast, weak systems with circulations (example: during the onset of a cyclone) are not adequately represented in the MSLP field as compared to the relative vorticity field at 850 hPa~\cite{lam1994use}. Therefore, relative vorticity is more suitable for tracking of cyclones at an early stage \cite{flaounas2014cyclotrack}, especially in cases of explosive cyclone-genesis compared to other variables. 

As the probability of detecting cyclones increases with increase in spatial resolution \cite{resolution}, and also that the relative vorticity, being a wind-based field, is sensitive to the spatial resolution of the data set \cite{flaounas2014cyclotrack}, we use a high spatial resolution of 0.5$^o$ $\times$ 0.5$^o$ for our analysis. For the analysis of the Seroja-Odette interaction, the spatial region of interest extends from 102$^o$ E to 125.5$^o$ E and from 5$^o$ S to 25$^o$ S (Fig.~\ref{fig_window} $a$). Similarly, in the case of the Noru-Kulap interaction, the study area chosen extends from 143$^o$ E to 169.5$^o$ E and from 23.5$^o$ N to 35.5$^o$ N (Fig.~\ref{fig_window} $b$). The selection of the spatial domain is made in a manner that ensures the elimination of any other neighbouring weaker cyclonic or anticyclonic vortices apart from the considered cyclone pair. So, inherently, we have made the assumption that the cyclone pair is not affected by the climate behavior outside the selected spatial region. Furthermore, in order to study the rapid intensification and weakening of the cyclones and the changes in their mutual interactions, we use a temporal resolution of 3 hours for the relative vorticity data set, as often used by cyclone track forecasters~\cite{ho2008cyclone,gupta2021complex}.

We perform our analyses to obtain the interaction structure of the two-dimensional relative vorticity field at the lower tropospheric level of 850 hPa, as commonly used for cyclone forecasts~\cite{Molinari1992,PILLAY2021100376}. Vorticity at 850 hPa has a stronger magnitude compared to vorticity at near surface heights (1000 hPa), especially for weaker circulations and therefore, is more robust when representing the strong upward motion of air. Hence, the 850 hPa relative vorticity field exhibits better continuity in the course of cyclone evolution~\cite{lam1994use} which is essential to deal with a CM event of two cyclones. Moreover, weaker cyclones have a shallow-lower tropospheric vertical depth (850-500 hPa) while the most intense cyclonic systems move with a deeper layer flow (850-200 hPa) \cite{neumann1979use,kimberlain2017tropical} which should be taken into account for producing optimal forecasts of cyclone tracks with the lowest mean forecast errors \cite{Velden1991}. Therefore, we also investigate the evolution of the network connectivity structure for higher tropospheric levels (500 hPa, 600 hPa, and 700 hPa), which not only allows us to verify the consistency of our results, but also to identify the transitions in the interaction structure of the binary cyclone system in the three-dimensional column of the atmosphere.

\subsection{ Construction of time-evolving directed networks}\label{subsec:2}

\begin{figure}[ht!]%
\centering
\includegraphics[width=0.46\textwidth]{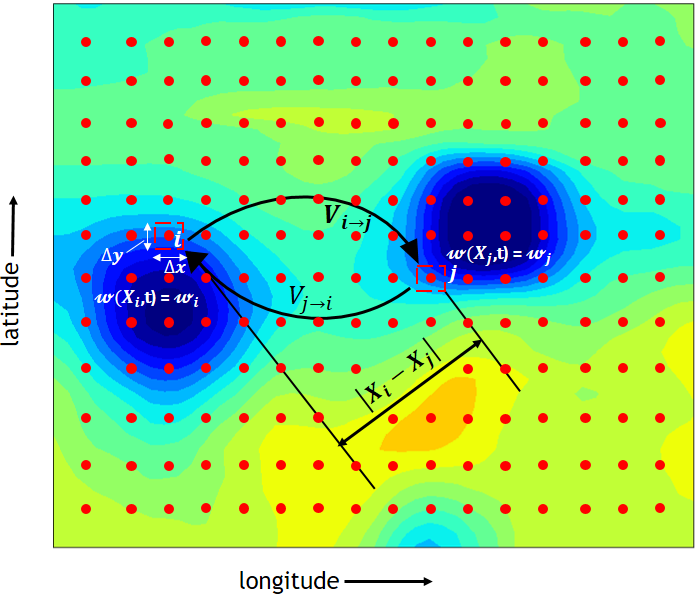}
\caption{Schematic illustration of the method of contruction of vorticity network for a binary cyclone system at a particular time step $t$. The solid red circles in the spatial domain represent the grid points or nodes. The velocity, induced by the flow element at node $i$ on node $j$ is shown in terms of $V_{i \rightarrow j}$. Similarly, $V_{j \rightarrow i}$ indicates the induced velocity on $i^{th}$ element by $j^{th}$ element. $\omega_{i}$ and $\omega_{j}$ represent the relative vorticities at $i^{th}$ and $j^{th}$ flow elements, respectively. The dashed square boxes represent the size of the fluid elements at the $i^{th}$ and $j^{th}$grid points.}
\label{fig1}
\end{figure}

The interactions between the different components of a complex system can be represented as a complex network in which each component can be considered as a node, and the pairwise interactions between the different components are represented as links between the nodes. Since its inception, the methods of complex networks have been the extensively followed holistic approaches to understand the collective behavior of many real-world complex systems~\cite{sujith2020complex, de2018network, avila2013classifying,chen2009cancer,davis2013global, boccaletti2006complex, abe2004scale, tsonis2004architecture,donges2009backbone,scarsoglio2013climate}.

We adopt the network approach to study the two-dimensional vortical interactions in binary cyclone system at a particular geopotential height in the current work. In this two-dimensional system, flow elements at grid points (Fig.~\ref{fig1}) are considered as nodes, and links between two nodes represent the interaction between these flow elements~\cite{taira2016network,krishnan2021suppression}. 
The magnitude of the velocity induced by the vorticity of a flow element at the $i^{th}$ grid point on another flow element at the $j^{th}$ grid point  ($V_{i \rightarrow j}$) (Fig.~\ref{fig1}) is estimated using the Biot-Savart law \cite{jackson1999classical} as,
\begin{equation}
\label{eqn:1}
V_{i \rightarrow j}={\frac{\mid \gamma_{i} \mid}{2\pi \mid X_i - X_j \mid}}
\end{equation}
where, $X_i$ and $X_j$ are the spatial location of the $i^{th}$ and $j^{th}$ grid point respectively. We take the absolute value of the circulation ($\gamma_{i}$ = $\omega (X_i) \Delta x \Delta y$) of the flow element at the grid point (node) $i$ as mentioned in 
Taira et al.~\cite{taira2016network}. Note that we are treating the spatial domain as planar (2D), as the extent of the domain is small enough (see Fig.~\ref{fig_window} $a$ and $b$). 
The Euclidean distance between the $i^{th}$ and $j^{th}$ nodes is represented by $\mid X_i - X_j \mid$. 
If the number of grids (nodes) in the flow domain is $N$, then the size of the induced velocity matrix is $N$$\times$$N$. The velocity induced by the flow element at the $i^{th}$ node on the element at the $j^{th}$ node ($V_{i \rightarrow j}$) is different from that induced by the element at the $j^{th}$ on the element at the $i^{th}$ node ($V_{j \rightarrow i}$) and therefore the matrix is asymmetric.

Further, following previous studies ~\cite{stolbova2014topology, gupta2021complex}, we consider only the highest 5$\%$ in the induced velocities to define the links in our network. This 95$^{th}$ percentile of the induced velocity is found to be the optimum choice for all our network calculations. 
Then, we build an adjacency matrix by registering the connections with links by 1. The rest of the elements of the adjacency matrix are filled by zeros. We also neglect self connections, i.e., the velocity induced by a flow element on itself is considered to be zero (Eq.~\ref{eqn:2}). Thus, we construct an unweighted directed network whose adjacency matrix $A_{ij}$ is represented as,    
\begin{equation}
\label{eqn:2}
A_{ij} = \begin{cases}
   1,& \text{if } i\neq j \text{ and }  V _{i \rightarrow j}>\text{threshold}\\
    0,              & \text{otherwise}
\end{cases}
\end{equation}
In this manner, we construct a time-varying spatial network from the vorticity field at every time instant to understand the evolution of the binary cyclone interaction.  

In relevance to the current study, Gupta et al. \cite{gupta2021complex} used a correlation based network spanning over a time window of 10 days, which encoded the interactions in the spatiotemporal field of MSLP data, to detect cyclone and their track in the basin. However, such a time-averaged network is unable to capture the evolution of the mutual interaction between two cyclones which occur over hourly to daily time scales. Therefore, instantaneous time-varying vorticity networks are a better alternative to not only detect cyclones but also to study their interaction with other cyclones.   

\subsection{Network measures}\label{subsec:3}

 In this analysis, we measure the strength of the nodes in the interacting flow domain through the network measure, degree \cite{barabasi2013network}, which counts the number of links or connections a node has with others. As our instantaneous vorticity network is a directed network, we distinguish the number of incoming and outgoing links to and from a node in terms of its \textit{in-degree} (${k^{in}_{i}}$) and \textit{out-degree} ($k^{out}_{i}$), respectively \cite{barabasi2013network}. 
$k^{in}_{i}$ is defined as,
\begin{equation}
\label{eqn:3}
{k^{in}_{i}}={\sum_{j=1}^{N} A_{ji}}
\end{equation}
and represents the in-degree of the flow element at $i^{th}$ node, where $i\neq j$. Through $k^{in}_{i}$, we can describe the impact of the induced velocities of the neighboring nodes at $i^{th}$ node in the interaction domain.

On the other hand, $k^{out}_{i}$ is defined as,
\begin{equation}
\label{eqn:4}
{k^{out}_{i}}={\sum_{j=1}^{N} A_{ij}}
\end{equation}
and represents the number of outgoing links from the flow element at the $i^{th}$ node, where $i\neq j$. $k^{out}_{i}$ can identify the strong vortices which induce velocities over a long distance in the interaction domain.   


\subsection{Separation distance between cyclones}\label{subsec:4}

The separation distance is a traditionally used metric to classify the interaction stages of binary cyclones~\cite{lander1993interaction,holland1993interaction} and the vortex merging process~\cite{cerretelli2003physical}. In the present study, the position of each cyclone is tracked on the basis of the geographical latitude and longitude of the center, obtained from  Weather Underground's Online database \cite{dist_cyclone}. We use the Haversine formula \cite{chopde2013landmark} to calculate the separation distance ($d$) between two cyclones. The steps used in this calculation is given in Appendix~\ref{sec:A11}.

\section{Results and Discussion }\label{sec:3}

First, we describe the evolution of the network connectivity structure of the two binary cyclones systems and try to relate it with the changes observed in their relative vorticity field (Sec.~\ref{subsec:5}). Thereafter, in Sec.~\ref{subsec:6}, we use the transitions obtained from the network-based parameters to classify the merging process into different stages.

\begin{figure*}[ht!]%
\centering
\includegraphics[width=0.90\textwidth]{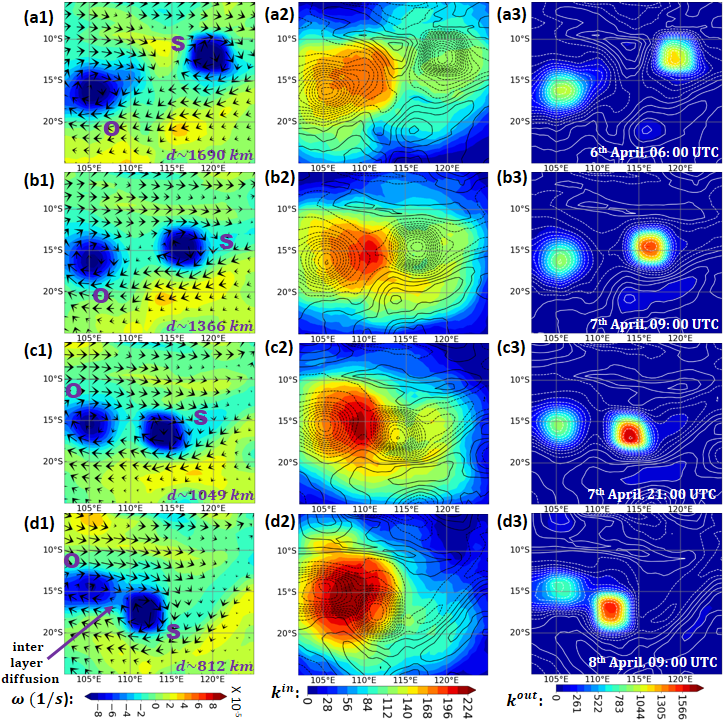}
\caption{The distributions of $\omega$ ($a1$-$d1$), $k^{in}$ ($a2$-$d2$), and $k^{out}$ ($a3$-$d3$) are presented during the interaction of Seroja ($S$) and Odette ($O$) at a geopotential height of 850 hPa. The time steps shown here are: ($a1$, $a2$, $\&$ $a3$) 6$^{th}$ April 2021, 06:00 UTC, ($b1$, $b2$, $\&$ $b3$) 7$^{th}$ April 2021, 09:00 UTC, ($c1$, $c2$ $\&$ $c3$) 7$^{th}$ April 2021, 21:00 UTC, and ($d1$, $d2$, $\&$ $d3$) 8$^{th}$ April 2021, 09:00 UTC. The velocity vector of the wind is shown in $a1$-$d1$. The vorticty contours of ($a1$-$d1$) are shown in the distributions of $k^{in}$ and $k^{out}$ for better understanding the changes of the interaction between two cyclones. Note that the negative $\omega$ is represented by the dotted line, while the portrayal of the vorticity contour by the solid line indicates the positive $\omega$. $k^{in}$ increases as the cyclones come closer by rotating around each other, while $k^{out}$ of the network can explain the loss or gain of the vorticity from each cyclone during the period. }\label{fig2}
\end{figure*}

\subsection{Degree analysis on vorticity network}\label{subsec:5}

\subsubsection{Seroja-Odette interaction}\label{subsubsec:1}

\begin{figure*}[ht!]%
\centering
\includegraphics[width=0.90\textwidth]{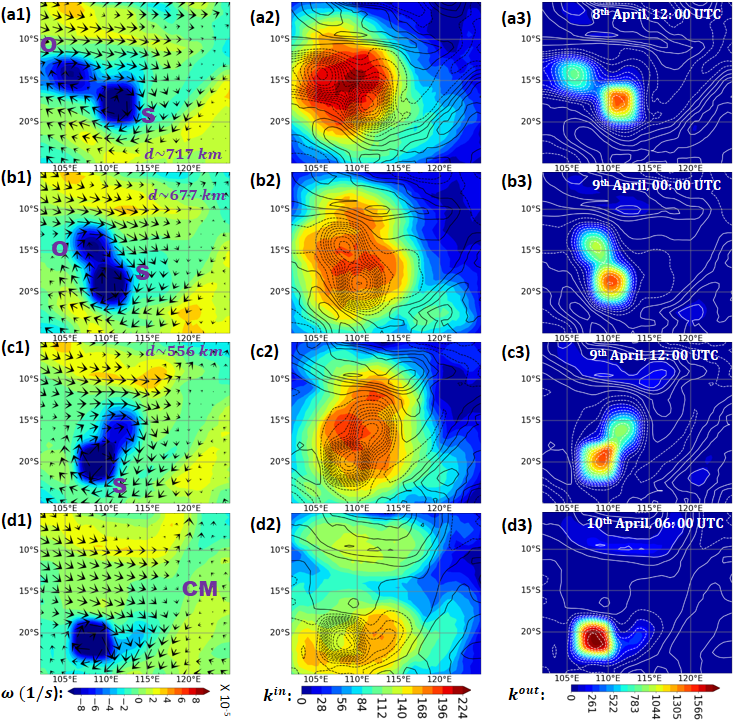}
\caption{The distributions of $\omega$ ($a1$-$d1$), $k^{in}$ ($a2$-$d2$), and $k^{out}$ ($a3$-$d3$) are presented during the interaction of Seroja ($S$) and Odette ($O$) prior to the CM at a geopotential height of 850 hPa. The time steps shown here are: ($a1$, $a2$, $\&$ $a3$) 8$^{th}$ April 2021, 12:00 UTC, ($b1$, $b2$, $\&$ $b3$) 9$^{th}$ April 2021, 00:00 UTC, ($c1$, $c2$, $\&$ $c3$) 9$^{th}$ April 2021, 12:00 UTC, and ($d1$, $d2$, $\&$ $d3$) 10$^{th}$ April 2021, 06:00 UTC. The velocity vector of the wind is shown in $a1$-$d1$. In this period, we can see a contraction of area covered by the higher $k^{in}$ as the interaction has a tendency to form a merged cyclone. During the CM event ($d1$), we can observe a higher $k^{out}$ at the center of merged cyclone ($d3$). The vorticty contours of ($a1$-$d1$) are shown in the distributions of $k^{in}$ and $k^{out}$.}\label{fig3}
\end{figure*}

We present the relative vorticity field at 850 hPa (Figs.~\ref{fig2} $a1$-$d1$), and the corresponding spatial distributions of $k^{in}$ (Figs.~\ref{fig2} $a2$-$d2$) and $k^{out}$ (Figs.~\ref{fig2} $a3$-$d3$), during the interval from 6$^{th}$ April 2021, 06:00 UTC to 8$^{th}$ April 2021, 09:00 UTC of the interacting period between Seroja and Odette in Fig.~\ref{fig2}. The strong negative values of $\omega$ (Figs.~\ref{fig2} $a1$-$d1$ and \ref{fig3} $a1$-$d1$) indicate a strong upward movement of air, causing the winds to rotate in a clockwise motion (as shown by the wind velocity vector), typical of cyclones in the southern hemisphere~\cite{southern_cyclone}. From Figs.~\ref{fig2} $a1$-$d1$, we find two distinct regions of negative $\omega$ values (blue color) in the vorticity field, indicating two cyclones.The vortex on the right side of the window represents cyclone Seroja (marked $S$)~\cite{khadami2021ocean}, whereas the vortex on the left side is the cyclone Odette (marked $O$). On 6$^{th}$ April at 06:00 UTC, these two cyclonic systems were $\sim$1690 km apart (Fig.~\ref{fig2} $a1$). Around this time, vorticity diffuses from the \textquotedblleft\textit{inner core}\textquotedblright (i.e., the intensified vorticity zone at the center of the cyclone) to the outer layers (i.e., the surroundings of the center) of the cyclones, dynamically changing the shape of the cyclones~\cite{leweke2016dynamics}. This phenomenon has been referred to as the intra-layer vorticity exchange~\cite{leweke2016dynamics}. 

Odette almost stays at the same location throughout the interacting period, from 6$^{th}$ April to 7$^{th}$ April 2021 (Figs.~\ref{fig2} $a1$-$c1$). In strong contrast, Seroja continuously moves towards Odette. 
As a consequence of this rapid movement of Seroja, $d$ significantly reduces (Figs.~\ref{fig2} $a1$-$c1$) during the interacting period. The detailed quantification of $d$ during this interaction is discussed later in Sec.~\ref{subsec:6}.

From the network connectivity structure, initially, we find that $k^{in}$ at the grid points near Odette is relatively larger than those near Seroja ( Fig.~\ref{fig2} $a2$) for a higher value of $d$. The higher $k^{in}$ near Odette denotes a higher vortical influence on the nodes of that regime by the other long-range or nearby nodes. On the other hand, $k^{out}$ is always observed to be higher inside the cyclones than the non-cyclonic regions in the spatial domain (Figs.~\ref{fig2} $a3$-$d3$), implying high outgoing links from the cyclones. As the vorticity diffusion occurs from the center to the outer layers of the cyclone and is limited to its outermost layer, we find a sudden drop of $k^{out}$ beyond a certain radius of the cyclone. Furthermore, the $k^{in}$ values are $\sim$ 10 times lower than the $k^{out}$ values which indicate that the number of links connecting both cyclones is comparatively less than the links arising from a cyclone. 

\begin{figure*}[ht!]%
\centering
\includegraphics[width=0.94\textwidth]{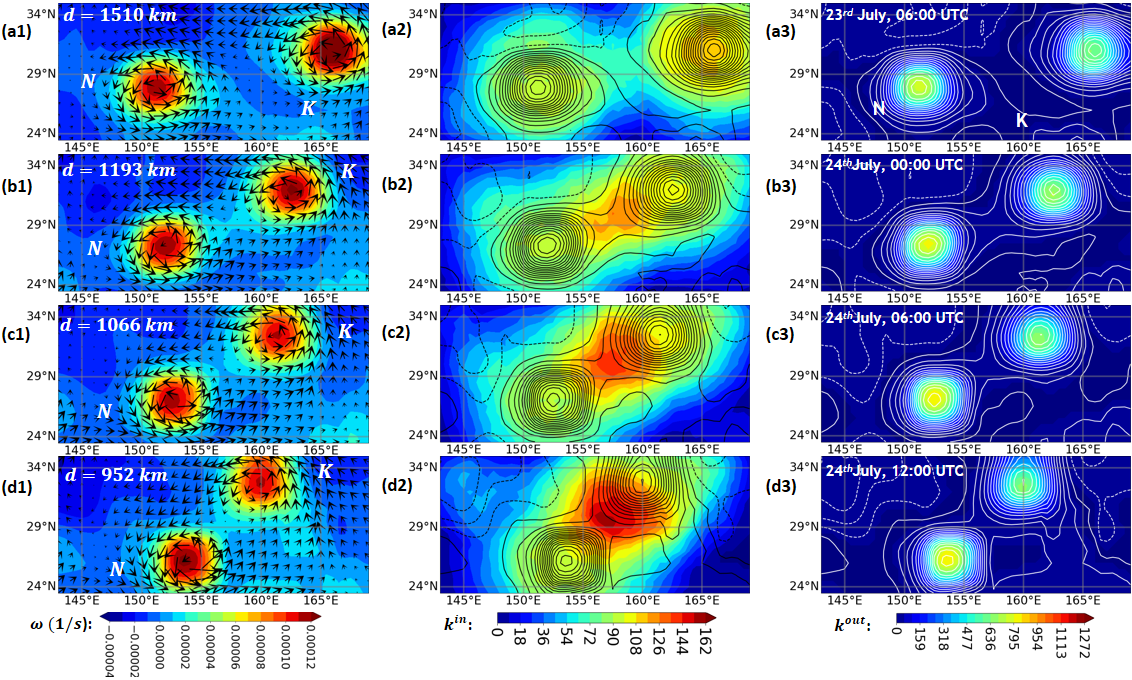}
\caption{The distributions of $\omega$ ($a1$-$d1$), $k^{in}$ ($a2$-$d2$), and $k^{out}$ ($a3$-$d3$) are presented during the interaction of Noru ($N$) and Kulap ($K$) at a geopotential height of 850 hPa during the period of 23$^{rd}$ July to 24$^{th}$ July 2017. The time steps shown here are: ($a1$, $a2$, $\&$ $a3$) 23$^{rd}$ July 2017, 06:00 UTC, ($b1$, $b2$, $\&$ $b3$) 24$^{th}$ July 2017, 00:00 UTC, ($c1$, $c2$, $\&$ $c3$) 24$^{th}$ July 2017, 06:00 UTC and ($d1$, $d2$, $\&$ $d3$) 24$^{th}$ July 2017, 12:00 UTC. The velocity vector of the wind is shown in $a1$-$d1$. The vorticty contours of ($a1$-$d1$) are shown in the distribution of $k^{in}$ and $k^{out}$ for better understanding the changes of Nor-Kulap interaction. Note that the positive vorticity is represented by the solid lines while dotted lines in vorticity contour indicate the negative vorticity. During the vorticity exchange between two cyclones, $k^{in}$ increases significantly between two cyclones. The higher $k^{out}$ at the center of Noru tells its strong impact on the neighboring nodes.}\label{fig4}
\end{figure*}

However, as $d$ reduces, the area covered by the higher $k^{in}$ nodes is observed to increase in between both cyclones (Figs.~\ref{fig2} $b2$-$d2$). After 2 days, when $d$ $\sim$ 812 km, the significantly higher $k^{in}$ between the cyclones (Fig.~\ref{fig2} $d2$), implies higher incoming links from surrounding regions thereby indicating that the vortical interactions occurring between both cyclones is very high. This high vorticity exchange \cite{cerretelli2003physical} occurs through the establishment of an inter-layer vorticity diffusion (Fig.~\ref{fig2} $d1$). Interestingly, compared to Seroja, Odette is closer to the higher $k^{in}$ area (see the contours of vorticity of the cyclones in Figs.~\ref{fig2} $b2$-$d2$) for lower values of $d$. In contrast, we find higher $k^{out}$ values at the grid points inside Seroja than Odette (Figs.~\ref{fig2} $a3$-$d3$). However, there is a slight drop in the magnitude of $k^{out}$ at the core of Seroja during inter-layer diffusion. All these observations suggest that Seroja exhibits more influence on the intermediate region than Odette during their interaction (Figs.~\ref{fig2} $a3$-$d3$). 

Further, we show the distributions of $\omega$ (Figs.~\ref{fig3} $a1$-$d1$), $k^{in}$ (Figs.~\ref{fig3} $a2$-$d2$), and $k^{out}$ (Figs.~\ref{fig3} $a3$-$d3$) of the Seroja-Odette interaction during the interval from 8$^{th}$ April 2021, 12:00 UTC to 10$^{th}$ April 2021, 06:00 UTC in Fig.~\ref{fig3}. After the establishment of inter-layer diffusion, the inner core of Odette moves towards Seroja (in Fig.~\ref{fig3} $b1$). Thus, we observe singly connected, dumbbell shaped cyclones \cite{cerretelli2003new} at this stage. Besides, during this stage of interaction, $k^{in}$ significantly shrinks within these singly connected cyclones (Fig.~\ref{fig3} $a2$), which signifies that the vortical influence from the cyclones on the interacting zone becomes lower compared to that found on 8$^{th}$ April 2021, 12:00 UTC (see Fig.~\ref{fig3} $a2$). On the other hand, the $k^{out}$  distribution of the cyclones (see Fig.~\ref{fig3} $b3$) bears a good resemblance to the corresponding vorticity distribution (Fig.~\ref{fig3} $b1$). At this time, $k^{out}$  of the core of Seroja intensifies further due to the intake of vorticity from Odette.

After 3 days, cyclone Odette decays, as indicated by the lower magnitude of negative $\omega$ (Fig.~\ref{fig3} $c1$). 
In the next stage ($d1$ in Fig.~\ref{fig3}), we see only a single vortex in the window, which confirms the occurrence of binary cyclone merging. The area covered by the higher $k^{in}$ is also observed to shrink simultaneously ($d2$ in Fig.~\ref{fig3}) during the CM event. Besides, we observe a higher $k^{out}$ at the center of cyclone Seroja, while the region of high $k^{out}$ abruptly vanishes around Odette ($d3$ in Fig.~\ref{fig3}).

\begin{figure*}[ht!]%
\centering
\includegraphics[width=0.94\textwidth]{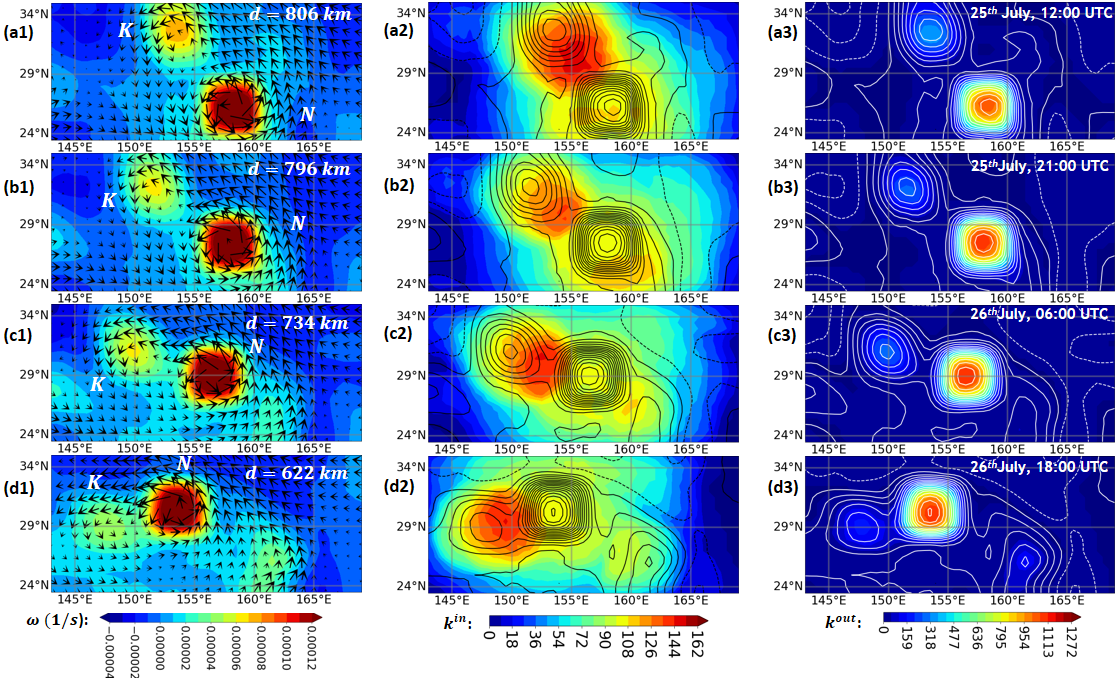}
\caption{The distributions of $\omega$ ($a1$-$d1$), $k^{in}$ ($a2$-$d2$), and $k^{out}$ ($a3$-$d3$) are presented during the interaction of Noru ($N$) and Kulap ($K$) prior to their merging at a geopotential height of 850 hPa during the period of 24$^{th}$ July to 25$^{th}$ July 2017. The time steps shown here are: ($a1$, $a2$, $\&$ $a3$) 25$^{th}$ July 2017, 12:00 UTC, ($b1$, $b2$, $\&$ $b3$) 25$^{th}$ July 2017, 21:00 UTC, ($c1$, $c2$, $\&$ $c3$) 26$^{th}$ July 2017, 06:00 UTC, and ($d1$, $d2$, $\&$ $d3$) 26$^{th}$ July 2017, 18:00 UTC. We can see closer proximity between two cyclones ($a1$-$d1$) before the merging. Besides, the reduction and increment of $k^{out}$ at the center of Kulap and Noru, respectively, helps to understand the merging after a vortcity exchange happens between them.The vorticty contours of ($a1$-$d1$) are shown in the distribution of $k^{in}$ and $k^{out}$.}\label{fig5}
\end{figure*}

Thus, we find the interesting spatiotemporal changes of the interacting field between Seroja and Odette in terms of $k^{in}$. Therefore, $k^{in}$ between the two cyclones may be taken as a quantitative measure of binary interaction through vorticity advection between the cyclones. In contrast, the dominating influence of Seroja over Odette is captured through $k^{out}$ at each time instant. The variation of $k^{out}$ within the cyclones and in the intermediate regime between them (for lower $d$) provides a perception of the vorticity diffusion, which might be a consequence of the induction of velocity of the flow elements due to their vorticity on the others.  

\subsubsection{Noru-Kulap interaction}\label{subsubsec:2}

Next, we investigate the binary interaction between Noru and Kulap, and the effect of neighboring air flows in Northwest Pacific~\cite{liou2019consecutive} during July 2017.  In contrast to the Seroja-Odette interaction, strong positive values in the relative vorticity distribution signify the rising motion of air causing the winds to be deflected counter-clockwise, as is typical for northern hemisphere cyclones~\cite{hoskins2002new} (Figs.~\ref{fig4} $a1$-$d1$ and ~\ref{fig5} $a1$-$d1$). During the period between 23$^{rd}$ July and 24$^{th}$ July 2017, Kulap is observed to change its track slightly towards the west, while a bit of eastward movement of Noru is seen \cite{li2021intensity}. After that, from 25$^{th}$ to 26$^{th}$ July 2017, a significant change in the direction of their movement results in reduction of $d$. Here, first, we discuss the interaction of these two cyclones during the period of 23$^{rd}$ July to 24$^{th}$ July 2017 (Fig.~\ref{fig4}). 

On 23$^{rd}$ July, Kulap and Noru are far apart from each other ($d$ $\sim$ 1510 km, Fig. \ref{fig4} $a1$). Similar to the Seroja-Odette interaction, at this stage, there is a slightly higher $k^{in}$ at the region closer to the center of Kulap compared to that of Noru, which signifies a dominating vortical influence from other regions on Kulap (see Fig.~\ref{fig4} $a2$). As also indicated from the higher $k^{out}$ values within Noru than Kulap (see the center of two cyclones in Fig.~\ref{fig4} $a3$), the vortical influence of Noru on Kulap dominates at this period. It is seen that, initially, $k^{in}$ is very low in the region between these two cyclones. As the cyclones rotate about each other, $k^{in}$ gradually increases in that region 
(see Figs.~\ref{fig4}~$b2$-$d2$). This increase in $k^{in}$ is again due to the vorticity exchange between the cyclones, which is prominently observed on 24$^{th}$ July 12:00 UTC (cf. Fig.~\ref{fig4} $d1$). On the other hand, the outer layers of Kulap facing Noru have comparatively higher $k^{in}$ than that of Noru facing Kulap which also signifies the higher impact of Noru on Kulap (see the higher $k^{in}$ between two cyclones in Figs.~\ref{fig4} $b2$-$d2$). Besides, the distributions of $k^{out}$ of the two cyclones highly resemble the vorticity distributions ($a3$-$d3$ of Fig. \ref{fig4}), as seen in Figs.~\ref{fig2} and \ref{fig3} ($a3$-$d3$) with the highest $k^{out}$ at the center of the cyclones. A significant decrease in $k^{out}$ is noticed as we move away from the center of the cyclone towards its outer layers. 

In addition, similar to Seroja-Odette interaction (Figs.~\ref{fig2} and \ref{fig3}), we find a sudden drop of $k^{out}$ outside a certain radius indicating the presence of higher interacting nodes inside the cyclones. 
Higher $k^{out}$ values at the nodes of Noru than those of Kulap within 23$^{rd}$ and 24$^{th}$ July corroborates the same understanding that the vortical influence of Noru highly dominates over that of Kulap on other nodes. Also, the vortical influence of the non-cyclone nodes has minimal effect compared to the cyclones, as seen from their near zero $k^{out}$ values. Thus, the higher $k^{in}$ during the inter-layer vorticity exchange between two cyclones confirms that the vortical influences at that zone mainly come from Noru. 

Furthermore, during the period 25$^{th}$-26$^{th}$ July 2017, Noru turns simultaneously to the north and then west while Kulap turns to the south-west \cite{li2021intensity} (Fig.~\ref{fig5} $a1$). Similar to the Seroja-Odette interaction, the vorticity core of Kulap is observed to diminish as the inter-layer vorticity interaction between the two leads to the formation of an unstable shape~\cite{cerretelli2003new} (Figs.~\ref{fig5} $b1$-$c1$). The corresponding $k^{in}$ distribution shows a significant shrinkage in the area covered by higher $k^{in}$ between both cyclones (comparing $b2$-$d2$ with $a2$ in the Fig.~\ref{fig5}). During this period, due to the closer proximity of Noru and Kulap, the interaction between them significantly reduces. A similar region of high $k^{in}$ but of relatively less magnitude is seen on the side of Noru opposite to that of Kulap (Fig.~\ref{fig5} b2), which subsequently  diminishes (Figs.~\ref{fig5} c2-d2). This additional $k^{in}$ region can be attributed to Noru's interaction with a neighboring weak vortex (at around 26$^o$ N, 162$^o$ E)  (Figs.~\ref{fig5}c1-d1), which is not of interest in our present work.  
On the other hand, a significant simultaneous reduction and increment of $k^{out}$ of Kulap and Noru, respectively, happen when $d$ $\sim$ 800 km (Fig. \ref{fig5} $a3$).  
During the merging, when Kulap moves towards Noru, $k^{out}$ at the location of Kulap reduces to almost zero (see Figs. \ref{fig5} $c31$-$d3$). However, from the establishment of inter-layer diffusion to the near CM event, we find a significant difference in $\omega$ and $k^{out}$ between these two cyclones, which is higher than that observed between Seroja and Odette. The increasing rate of vorticity absorption of Noru from Kulap during this phase is the primary reason for that.    

Nevertheless, the topology of the interaction between the cyclones in Secs. \ref{subsubsec:1} and \ref{subsubsec:2}, as the cyclones in each pair merge, are found to be almost similar, although they occur in different cyclone basins in opposite hemispheres. However, from our spatiotemporal analysis in Figs.~\ref{fig2}-\ref{fig5}, we can infer a few notable pieces of information:

(i) As $k^{out}$ is high only over the cyclones, it indicates that the high incoming links in the region between two cyclones also come out from the cyclones, indicating a high vorticity interaction between both cyclones. 

(ii) $k^{out}$ values are $\sim$10 times larger than $k^{in}$ in the interaction of binary cyclones. This higher magnitude of $k^{out}$ in comparison with $k^{in}$ indicates stronger interactions within the cyclone and much less interaction with nodes farther than a certain distance. However, the region of interaction in between the two cyclones have vortical connections primarily with nodes within the cyclones, and is dominated by the cyclone having higher $k^{out}$. 

(iii) A sharp decline of $k^{out}$ outside a certain radius of the cyclones is indicative of grouping tendencies of the cyclone nodes \cite{gupta2021complex}
within the network. 

(iv) While $k^{out}$ helps cyclones to be easily identifiable in the network topology, beyond a certain separation distance, $k^{in}$ can be a quantitative measure of binary interaction between the two cyclones.
   
Thus, we can reveal the interaction dynamics of the binary cyclone system by understanding the induction of velocity by one flow element on the others in the spatial domain. In the next section (Sec.~\ref{subsec:6}), we test the performance of induced velocity-based network indicators in quantifying the dynamical transitions during a binary cyclone interaction leading to a CM.

\subsection{Identification of interaction stages leading to cyclone merger}\label{subsec:6}

\begin{figure*}[ht!]%
\centering
\includegraphics[width=1.0\textwidth]{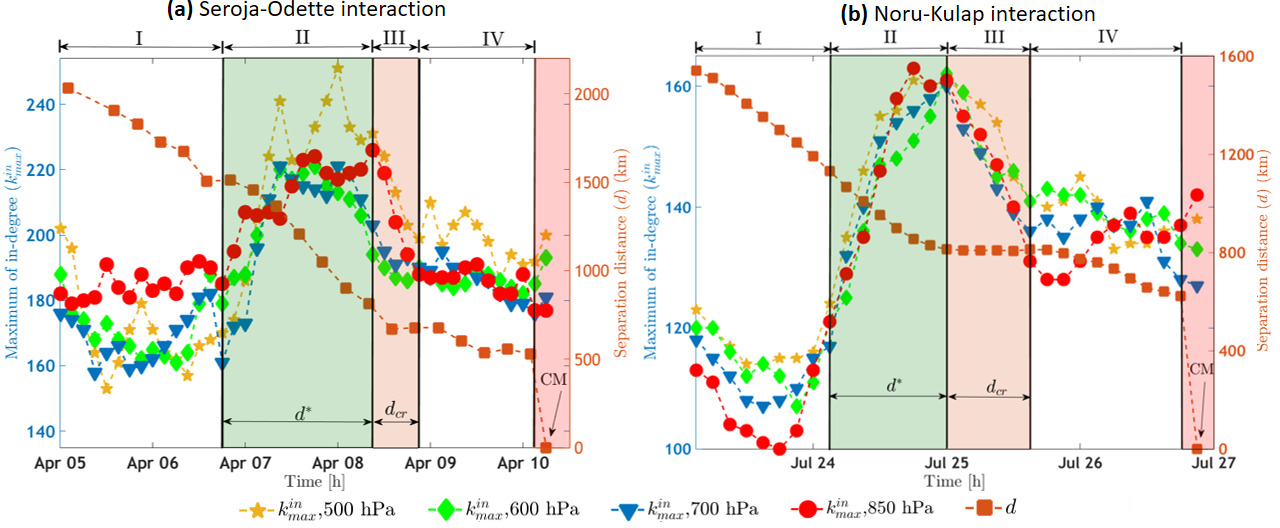}
\caption{We find the maximum value of in-degree ($k^{in}_{max}$) of the network with time for (a) Seroja-Odette and (b) Noru-Kulap interactions at different geopotential heights (500 hPa, 600 hPa, and 850 hPa). Here, we classify the binary cyclone interaction into four stages based on the variation of $k^{in}_{max}$ at 850 hPa. During the interaction, increment and reduction of ($k^{in}_{max}$) indicate strong and weak interaction between two cyclones, respectively.}\label{fig6}
\end{figure*}

To classify the merging process into different stages, we further quantify the transitions found in the spatial distributions of the network measures by computing the maximum of $k^{in}$ ($k^{in}_{max}$). To be more specific, the changes of $k^{in}_{max}$ seen in the nodes located at the intermediate zone between the two cyclones or the outer layers of the weaker cyclone facing the stronger cyclone (as seen in Figs.~\ref{fig2}-\ref{fig5} $a2$-$d2$) enable us to characterize the transitions during the binary cyclonic interaction. We use the variation of $k^{in}_{max}$ during the interaction period of the two binary cyclone systems (Fig.~\ref{fig6}). 

First we consider the variation of $k^{in}_{max}$ for the interaction between cyclones, Seroja and Odette at different geopotential heights (Fig.~\ref{fig6}$a$). We classify the interaction period of this binary interaction based on the variation of $k^{in}_{max}$ at 850 hPa since the height is conventionally used by cyclone forecasters ~\cite{Molinari1992,PILLAY2021100376}. During the interaction period, we find a stage-$I$, when $k^{in}_{max}$ seems to be more or less invariant from 5$^{th}$ April 2021, 00:00 UTC to 6$^{th}$ April 2021, 18:00 UTC (where, $d$ reduces from 2034 km to 1512 km). In this time period, $d$ is so large that the interaction between the two cyclones is not significant and remains independent of the mutual vorticity transport. However, a significant increment in $k^{in}_{max}$ is seen during the period corresponding to reduction in $d$ from 1455 km to 812 km, which signifies the presence of a stronger interaction between both cyclones as they come closer. We regard this interaction period (from 6$^{th}$ April 2021, 21:00 UTC to 8$^{th}$ April 2021, 09:00 UTC) as stage-$II$. So far, we have shown that the transitions in the mutual interaction at these two stages can be more clearly distinguished from $k^{in}_{max}$, although we find minor difference in the reduction rate of $d$ between the stages - $I$ and $II$. 

Further, after reaching its maximum value, we observe a sharp fall in $k^{in}_{max}$ from 8$^{th}$ April 2021, 09:00 UTC to 8$^{th}$ April 2021, 21:00 UTC. We regard this period as stage-$III$. Throughout this stage, $d$ is almost constant. After that, close to the merging of Odette into Seroja, we find nearly constant values of $k^{in}_{max}$ from 8$^{th}$ April 2021, 21:00 UTC to 10$^{th}$ April 2021, 03:00 UTC. We regard this period as stage-$IV$. At this stage, the constantly low $k^{in}_{max}$ denotes that there are no longer interactions between the cyclones since they approach CM. We observe a similar trend in $k^{in}_{max}$ for this binary interaction at other geopotential heights (Fig.~\ref{fig6} $a$). 

Hence, we find four distinct stages of the Seroja-Odette interaction before a CM event. In this context, a previous study~\cite{couder1986experimental} discussed a spontaneous formation of the coupling between two vortices of the opposite sign. However, this study primarily focused on the elementary processes of vortex pairing. Since then, a number of numerical studies~\cite{brandt2007physics,josserand2007merging} have focused on the vortex pairing and merging based on their separation distance. Recently, Cerretelli \& Williamson~\cite{cerretelli2003physical} and Josser \& Rossi~\cite{josserand2007merging} showed different diffusion-convection stages for vortex merging based on the separation distance through experiments and numerical simulations, respectively. They found three phases before a merged diffusion (or complete merging), which are: a first diffusion (where separation slowly reduces), convection (separation reduces rapidly), and a second diffusion. However, as the changes in the interaction between two cyclones (Fig.~\ref{fig6}~$a$) are not consistent with that of $d$, as shown in the present study, $d$ may not be the most suitable measure to classify the phases of a binary cyclone interaction and merger. Moreover, these model-based experimental investigations did not take the large-scale environmental circulations into consideration, which is one of the crucial factors in the interaction of binary cyclones.     

Next, we check the variation of $k^{in}_{max}$ for the Noru-Kulap interaction. Similar to Seroja-Odette, for the Noru-Kulap interaction (Fig.~\ref{fig6}~$b$), we do not find a significant change of $k^{in}_{max}$ when $d$ reduces from 1540 km to 1150 km (stage-$I$). A rapid increase in $k^{in}_{max}$ can be found when $d$ becomes approximately 1000 km. However, the increment of the measure is continuously observed till the time when $d$ $\sim$ 812 km. The range of $d$ for which $k^{in}_{max}$ increases is 1150-812 km (stage-$II$). Interestingly, we can identify two distinct types of behavior of $k^{in}_{max}$ at these two stages, although the reduction rate of $d$ is the same. Again, we see a reduction and a saturation of $k^{in}_{max}$ in stages-$III$ and $IV$, respectively. Thus, we can identify four stages for a binary cyclones interaction before the CM happens. Further, the behavior of the interactions at different stages of these two chosen binary cyclone systems is encapsulated in Table~\ref{Tab:1}.

However, an increasing trend of $k^{in}_{max}$ in stage-$II$ (where the threshold $d$ = $d^{*}$) of the binary interaction denotes that $k^{in}_{max}$ is a promising tool to provide an idea of the particular separation distance beyond which the Fujiwhara interaction comes into play as well as give an early indication of a CM event. Furthermore, we may select a critical range of separation distance ($d_{cr}$), when $k^{in}_{max}$ starts to reduce (seen at the stage-$III$). After a sharp fall of $k^{in}_{max}$ in stage-$III$, a constant behavior before the merging increases the significance of $d_{cr}$. However, to be on the safe side, we must follow the trend of $k^{in}_{max}$ in stage-$II$ to issue the awareness of the cyclone merging. Previously, a large number of studies \cite{liu2015interactions, liou2016generalized} defined a threshold distance to decide whether two cyclones start to interact or not, and a separation distance within 1050 to 2250 km was found as the critical value for interactions of cyclones. However, estimating the separation distance to get an early indication the cyclone merger based on vorticity network-based measures proposed in the present study has a strong potential for a substantially improved forecast accuracy.

\begin{figure*}[ht!]%
\centering
\includegraphics[width=1\textwidth]{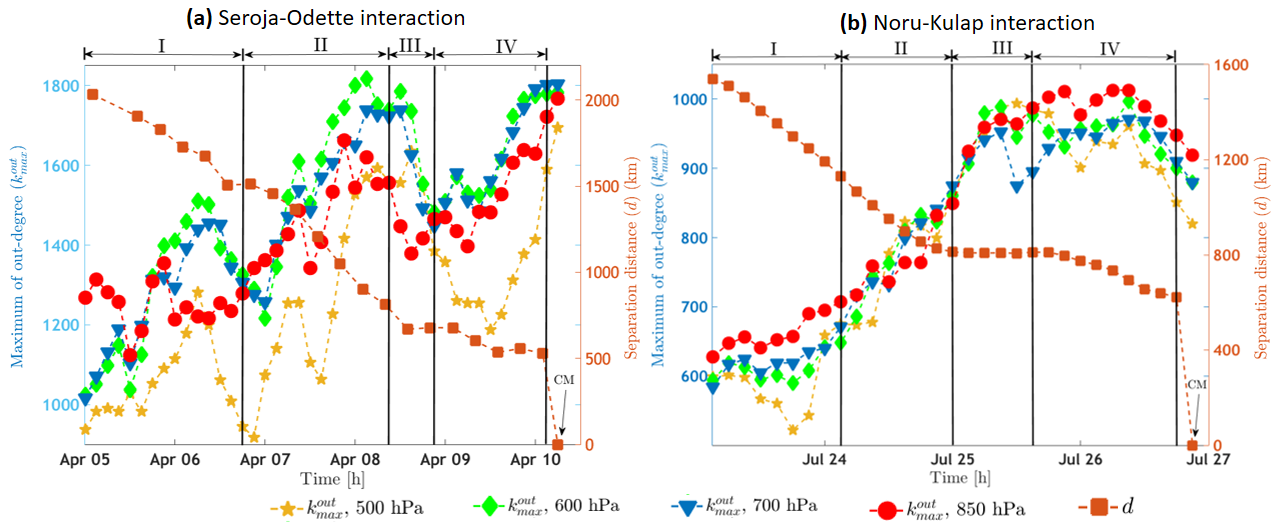}
\caption{The variation of maximum out-degree ($k^{out}_{max}$) of Seroja-Odette interaction ($a$) and Noru-Kulap interaction ($b$) is shown at different stages ($I$, $II$, $III$, and $IV$). The changes in $k^{out}_{max}$ at each time instant helps to quantify the variations of strength of the dominating cyclone. The significantly high $k^{out}_{max}$ in stage-$IV$ indicates the impact of the merged cyclone (in $a$). In contrast, we see a little drop of $k^{out}_{max}$ of the merged cyclone for Noru-kulap interaction ($b$).}\label{fig7}
\end{figure*}

In addition, we also consider the maximum $k^{out}$ ($k^{out}_{max}$) to understand the changes in the influence of the dominating cyclone at the different interaction stages (Fig.~\ref{fig7}).
The stages obtained based on $k^{in}_{max}$ are kept the same for the analysis of $k^{out}_{max}$. Let us consider the finding of $k^{out}_{max}$ for the Seroja-Odette interaction (Fig.~\ref{fig7}a). During stage-$I$, $k^{out}_{max}$ increases for sometime, although it saturates later. However, after $d$ $\sim$ 1512 km, as the cyclones come closer (stage-$II$), $k^{out}_{max}$ becomes high, which signifies that the vortical influence of the dominating cyclone continuously increases in the interacting field. 
On the other hand, when the cyclones reach stage-$III$ (in Fig.~\ref{fig7} $a$), a slight drop in $k^{out}_{max}$ happens during the inter-layer vorticity exchange (Fig.~\ref{fig2} $d3$). Thus, the increment and the sudden drop of $k^{out}_{max}$ in stages-$II$ and $III$ reveal the convective and diffusive nature of the interaction, respectively.
In stage-$IV$, the increase of $k^{out}_{max}$ indicates the strength of Seroja by which Odette is being absorbed.  During the CM event, a significantly high $k^{out}_{max}$ is related to the strength of the merged cyclone, making it more dominant in the influence over the network.  

\begin{table*} 
 \caption{Details of interaction for the two binary cyclone systems, Seroja-Odette and Noru-Kulap, are summarized here.}\label{Tab:1}
\centering
\setlength\arrayrulewidth{1.25pt}
\begin{tabular}{|p{2.3cm}|p{2.3cm}|p{2.3cm}|p{2.3cm}|p{2.3cm}|p{2.3cm}|}
\hline
\begin{flushleft}Binary cyclone interaction\end{flushleft}&\begin{flushleft}Factors\end{flushleft} & \begin{flushleft}Stage-$I$\end{flushleft} & \begin{flushleft} Stage-$II$\end{flushleft} & \begin{flushleft} Stage-$III$\end{flushleft} & \begin{flushleft} Stage-$IV$\end{flushleft}\\
\hline
& $d$ (km) & $\sim$ 1512-2033 & $\sim$ 812-1512 & $\sim$ 669-812 & $\sim$ 527-668 \\
\cline{2-6}
 & interaction between cyclones ($k^{in}_{max}$)& nearly constant & significantly increases & reduces & nearly constant                 \\
\cline{2-6}
Seroja-Odette &impact of dominating cyclone ($k^{out}_{max}$)   &nearly constant& increases   &slightly reduces    &again increases till CM                   \\
\cline{2-6}
&dominating cyclone   &Seroja   &Seroja    &Seroja &Seroja                   \\
\hline
&$d$ (km)   &$\sim$1133-1540 & $\sim$812-1131    &$\sim$806-812   &$\sim$622-805                \\
\cline{2-6}
&interaction between cyclones ($k^{in}_{max}$)   &nearly constant& significantly increases   &reduces    &nearly constant                 \\
\cline{2-6}
Noru-Kulap &impact of dominating cyclone ($k^{out}_{max}$)   &nearly constant& significantly increases   &saturated after increasing    &slightly reduces prior to CM\\
\cline{2-6}
&dominating cyclone   &Noru   &Noru    &Noru &Noru                   \\
\hline
\end{tabular}
\end{table*}

In contrast to the Seroja-Odette interaction, we find a rapid increment of $k^{out}_{max}$ in the stages -$II$ and $III$ of Noru-Kulap interaction (in Fig.~\ref{fig7}~$b$) due to the increasing rate of vorticity absorption of the dominating cyclone (here, Noru). Also, we do not see any appreciable drop in $k^{out}_{max}$ during the inter-layer diffusion between Noru and Kulap. The significant vorticity diffusion from Kulap into Noru till stage-$III$ may be the probable reason behind this continuous increment in the measure. However, we find a slight drop in $k^{out}_{max}$ in stage-$IV$ (close to the CM) probably due to the presence of a neighboring weak vortex~\cite{liou2019consecutive}. Considering the overall trend, $k^{out}_{max}$ seems to vary almost proportionally with $k^{in}_{max}$ in the stages-$I$, $II$, and $III$ for the Seroja-Odette interaction and in the stages-$I$, $II$ for the Noru-Kulap interaction. For an easier understanding of the behavior of the binary cyclone systems before a CM, we further epitomize the information obtained from the analysis of $k^{out}_{max}$ in Table~\ref{Tab:1}.  

To summarize, the early increment in stage-$II$ (Figs.~\ref{fig6} and~\ref{fig7}) makes both $k^{in}_{max}$ and $k^{out}_{max}$ promising candidates for providing vorticity interaction-based early warning signals of the CM in binary cyclone systems, although significant differences are seen in the stages-$III$ and $IV$ of $k^{out}_{max}$ between the interaction systems chosen from both hemispheres. On the other hand, quantification of $k^{out}_{max}$ seems to be helpful for understanding the dynamic changes of the dominating cyclone in a better way. Thus, adopting an unweighted directed network on the relative vorticity data provides a clear perception of the transitions in the binary cyclone merging process and helps forecast the merging event.  

\section{Conclusion}\label{sec:4}

In this study, we try to understand the underlying dynamics of the interaction and merging of two cyclones. To better comprehend the changes in the connectivity structure during the interaction between two cyclones at their proximity, we adopt an innovative network approach based on the pairwise induced vorticity interactions among the flow elements. Following this framework, we perform a degree analysis of the constructed time-evolving directed induced velocity networks. The in-degree of the vorticity network shows a simultaneous increasing and decreasing trend 
before and after the inter-layer diffusion between two cyclones. On the other hand, the higher out-degree of the cyclones signifies that the zone of interaction between the cyclones is mainly influenced by the dominating cyclone of the binary cyclone system. Thus, using the distributions of in-degree, we can understand the dynamics of interaction, while the distributions of out-degree help to identify the stronger cyclone during each time step of the interaction until the CM. 

Further, a rapid fall in out-degree, observed after a certain distance from the outside of each cyclone, indicates the occurrence of stronger interaction within the cyclone, and thus, the distribution of out-degree can clearly identify the cyclone. The changes in the out-degree provide an insight into the vorticity interaction that is dominated by the cyclonic regions. However, the vorticity diffusion, which may be a consequence of the velocity induced by the flow elements on the others, can be further studied to have a deeper understanding.  
It is noteworthy from the present study that we can classify the transitions of the binary cyclone interaction into four stages before CM occurs based on the quantification of maximum in-degree. Furthermore, an early growing trend of maximum in-degree and maximum out-degree in stage-$II$ helps to get an awareness of the occurrence of binary cyclone merging events.  

Thus, the complex network representation of the spatiotemporal relative vorticity field enables us to directly study the interaction structure of the vorticity field, making it a very suitable approach to gain incisive insights into the interaction process of binary cyclones. The method could be further applied to study different types of cyclone interactions, such as partial merger, partial straining out, and elastic interaction in different cyclone basins. The study of the differences in the interaction structure between co-rotating and counter-rotating (such as cross-equatorial twin cyclones) cyclone pairs could also be outlined as one of the future scopes of this work. Furthermore, combining this network approach with the physics-inspired machine learning algorithms can also be used to obtain a deeper understanding of the sudden track changes of cyclones caused due to the interaction of the cyclone with large-scale low-level cyclonic vortices such as the monsoon gyre. Such a detailed characterization of the connectivity structure of the different types of binary cyclones interactions is an essential step towards improving cyclone track forecasts.

\begin{acknowledgments}
R. I. S acknowledges the funding provided by the Office of Naval Research Global (Grant No. N62909-22-1-2011, contract monitor: Dr. Philip Chu). S.D. would also like to acknowledge the Post Doctoral Researcher fellowship under the same grant at Indian Institute of Technology Madras. S.G. and J.K. are supported by the Climate Advanced Forecasting of sub-seasonal Extremes (CAFE) project which has received funding from the European Union’s Horizon 2020 research and innovation programme under the Marie Skłodowska Curie Grant Agreement No. 813844. N.M. acknowledges the BMBF grant ClimXtreme (No. 01LP1902J) "Spatial synchronization patterns of heavy precipitation events".
\end{acknowledgments}

\section*{Data Availability Statement}

The data/reanalysis that supports the findings of this study are publicly available online: ERA5 Reanalysis data~\cite{hersbach2020era5}, https://cds.climate.copernicus.eu/, last accessed on 16 Dec 2021. 
For tracking the coordinates of the cyclone, we use Weather  Underground’s  Online  database~\cite{dist_cyclone}, https://www.wunderground.com/hurricane, last accessed on 20 Dec 2021.

\section*{Declarations: Conflict of interests}
The authors declare no competing interests.

\appendix
\section{Calculation steps for estimating the separation distance}\label{sec:A11}

Following Eqs. (\ref{eqn:6}-\ref{eqn:8}) are followed to calculate the separation distance between two nearby cyclones in the present work.

\begin{equation}
\label{eqn:6}
B_1=\sin^2{\frac{\delta \phi}{2}}+\cos{\phi_{1}}*\cos{\phi_{2}}*\sin^2{\frac{\delta \theta}{2}}
\end{equation}

Here, $\phi_1$ and $\phi_2$ are the latitudes of two cyclones at a particular time instance. We calculate the difference in latitude, $\delta \phi = \phi_1 - \phi_2$. Similarly, $\delta \theta$ is the difference in longitude of two corresponding cyclones.

\begin{equation}
\label{eqn:7}
B_2=2\tan^{-1}(\sqrt{B1}, \sqrt{1-B1}) 
\end{equation}
\\
The separation distance between two cyclones can be calculated as,
\begin{equation}
\label{eqn:8}
d=R*B_2
\end{equation}

\bibliography{aipsamp.bib}

\end{document}